# Enhancing Strategic Information Security Management in Organizations through Information Warfare Practices

*Research in Progress*

**Abid Hussain Shah**
Department of Computing and Information Systems
The University of Melbourne
Victoria, Australia
Email: abidh@student.unimelb.edu.au

**Atif Ahmad**
Department of Computing and Information Systems
The University of Melbourne
Victoria, Australia
Email: atif@unimelb.edu.au

**Sean B. Maynard**
Department of Computing and Information Systems
The University of Melbourne
Victoria, Australia
Email: sean.maynard@unimelb.edu.au

**Humza Naseer**
Department of Computing and Information Systems
The University of Melbourne
Victoria, Australia
Email: humza.naseer@unimelb.edu.au

## Abstract

Today's organizations use control-centred security management systems as a preventative shield against a broad spectrum of attacks. However, these have proven to be less effective against the customized and innovative strategies and operational techniques used by Advanced Persistent Threats (APTs). In this short paper we argue that to combat APTs, organizations need a strategic-level shift away from a traditional prevention-cantered approach to that of a response-cantered one. Drawing on the information warfare (IW) paradigm in military studies, and using Dynamic Capability Theory (DCT), this research examines the applicability of IW capabilities in the corporate domain. We propose a research framework to argue that conventional prevention-centred response capabilities; such as incident response capabilities and IW-centred security capabilities can be integrated into IW-enabled dynamic response capabilities that improve enterprise security performance.

# 1   Introduction

Organizations face a large variety of attacks against their information systems (IS), threatening to breach existing security management practices. Traditionally, the approach used by the organizations to counter these threats is control-centred and preventive in nature (Baskerville 2014; Ahmad et al. 2014). This approach is generally successful against known or predicted threats, however, its efficacy against the skilled, customized, innovative strategies and improvised techniques used by threats such as Advanced Persistent Threats (APTs), is marginal (Ahmad et al., 2019). This research adopts the arguments of Baskerville et al. (2014) which suggests that to deal with APTs, organizations need a strategic-level shift away from traditional prevention techniques. Drawing on the concepts of Warfare Strategies (non-kinetic in nature only; not involving physical fights) and using Dynamic Capability (DC) Theory, this research examines the applicability of Warfare-enabled Dynamic Incident Response Capabilities (WEDIRC) in the corporate domain for enhancement of the security performance.

Ensuring the protection of information resources has become a costly, time consuming task for organisations and is gaining increased recognition (Baskerville 2005; Ahmad et al. 2012; Wilson and Warkentin 2013; Park et al. 2012). An information security (InfoSec) threat or incident is an adverse event such as violations of policy and/or unauthorized access (Ahmad et al., 2014). Security threats may be Incidental (human error, technical failure or forces of nature affecting security) or Purposive (intentional breaches to an organisation's security driven by human intelligence) threats (Ahmad et al. 2014). Since purposive threats are becoming increasingly disciplined and strategic, they make the security environment uncertain and more challenging (Lemay et al. 2018). This research focuses on purposive threats.

Since the context of InfoSec is changing swiftly and security threats are becoming more complex, incidents are recurring more persistently. The situation suggests that a paradigm shift is required to enhance the capabilities of organizations (Baskerville 2005, 2014; Denning & Denning 2010). Therefore, we infer that the organizational response capability is inadequate to counteract the capabilities of APTs. Information Warfare (IW) strategies are inherently more response-oriented and better suited to address the challenge of adapting to a complex and evolving threat landscape (Baskerville 2014; Ahmad et al. 2014; Joint Chiefs of Staff. 2003; Field Manual 3-05.30 2005). We propose that to enhance information system security, a paradigm shift is needed, towards the IW paradigm. Literature suggests that not much is known about how organizations can implement a response-centred approach using the IW paradigm. To address this gap, this research explores the following question: *"How can appropriate information warfare strategies improve the organizational security response to security threats."*

To answer the question, our research proposes a IW-enabled organizational response framework. Although almost all organizations deal with external and internal threats to their IS, our research caters for organizations (medium and large sized organizations) for which information systems play a core role in the value creation process. This paper comprises four major sections. In the next section we introduce the theoretical framing for the paper. We then review the extant literature on security incident response and the capabilities of IW. This leads to a discussion on IW Enabled Dynamic IR Capabilities. Following this we introduce a conceptual framework of IW enabled security response before offering a set of propositions and concluding the paper.

# 2   Theoretical Framing – Dynamic Capabilities (DC) Theory

There has been a substantial focus on the conceptualization of dynamic capabilities, primarily considering rapidly changing technological advancements. Teece et al. (1997), proposes the DC approach as an extension of the resource-based view (RBV). RBV is static in its nature and inadequate in frequently changing environments (Priem 2007). Teece et al. (1997) addresses this aspect by defining DC as "the firm's ability to integrate, build, and reconfigure internal and external competences to combat rapidly changing environments." For reconfiguration of competencies, strategic process management is essential. DC are the organizational and strategic routines by which firms achieve new resource configurations in changing environments (Eisenhardt et al. 2000). To maintain a competitive advantage, it is necessary to exploit existing opportunities, create opportunities and remain viable for looming threats. DC are the capacity (a) to sense and shape opportunities and threats, (b) to seize the opportunities, and (c) to maintain competitiveness (Teece 2007). The literature provides insights for incident learning activities (Levitt & March 1988; Shedden et al. 2010). Naseer (2018) argues that organizations can manage incidents response efficiently following a DC approach. For this purpose, organizations need to adapt to the changing threat landscape. Better adaptation occurs when an interdisciplinary approach is followed (Naseer et al. 2017). By incorporating practices of other disciplines, an integrative and dynamic phenomenon occurs which harbours more opportunities for capturing changing environments avoiding introspective and parochial views (March 1991; Dodgson 1993). IW capabilities are response oriented and can be integrated to build DC resulting in enhanced security performance. We argue that DC theory is appropriate for this research as the threats faced by organisations evolve swiftly, create high levels of uncertainty and the concept of building new competencies and integrating them into current practice (incorporating IW capabilities with existing IR capabilities) is the focus of this research.

# 3 Literature Review

To establish the background of this research and to create a firm foundation for advancing knowledge, a literature review was conducted following Webster and Watson (2002). Articles published in IS, IW and information management journals and conferences were consulted based on key words and phrases such as; 'information security strategy, management', 'advanced persistent threats', 'IS controls', 'warfare strategies', 'dynamic capabilities', 'incident response' and 'information security risk management'. Initially, 124 articles were identified in addition to 10 books and 5 field manuals. The focus was narrowed by reading article abstracts and removing those not specifically in the area. Further, we read the articles, concentrating on research questions and outcomes, to identify those articles that focus on improving InfoSec response and what can be adopted from IW strategies. This process resulted in 53 articles. Using forward and backward chaining through the references of these articles, 15 additional articles were identified (totalling 68 articles, two books and 4 field manuals of United States military on relevant IW capabilities). Through the analysis of the literature, our proposed conceptual framework (Figure 1) was developed. We have kept a DC theory focus throughout to keep the research within the correct theoretical bounds. The review revealed that incident response capabilities are largely prevention oriented whilst IW capabilities are more response oriented (Baskerville 2005; Baskerville et al. 2014; Ahmad et al. 2014).

## 3.1 Incident Response (IR) Capabilities

Purposive threats are a significant challenge to the security of organizations as they are organized, strategic, persistent and evolving (Lemay et al., 2018). They focus on discovering vulnerabilities in organisational defences. In the InfoSec paradigm, threats can be handled by applying controls, including: (1) Formal (e.g. policy and risk management) (2) Informal (e.g. training), and (3) Technological controls (e.g. firewalls, anti-virus software, and intrusion detection systems) (Sveen et al. 2009; Whitman and Mattord 2017). Since threats are treated by controls, the process of security management emerges to be a control-centred process (Baskerville 2014). To appropriately define the IR capabilities, we have taken analogy from the DC theory perspective. In DC theory the term 'capabilities' refers to "the critical role of strategic management in appropriately 'adapting, integrating, and reconfiguring internal and external organizational skills, resources, and functional competences to match the requirements of changing environment" (Teece et al. 1997). For this research we have defined Incident Response Capabilities as: ***IR capabilities comprise of all available controls (formal, informal and technical), practices and processes that address security threats to IS in performing IR functions***. Another perspective related to the capabilities, which is emerging in IS literature, discusses peculiarities of skill sets of persons involved, as well as the processes and the technologies used for responding to incidents (Werlinger et al. 2010).

While organizations are finding it increasingly difficult to ensure the protection of their information resources, current 'best practice' InfoSec response assumes a prevention-oriented strategy (Shedden et al. 2010; Ahmad et al. 2014; Baskerville 2005, Baskerville et al. 2014). This practice has worked well for known threats. However, the fact remains that despite employing a range of preventative controls, incidents are still occurring frequently. Although many organizations retain incident response teams to address incidents, maintaining standing IR teams is not cost-effective (Shedden et al 2010; Hove et al. 2014; Ahmad et al. 2014). Therefore, only a few organizations (financial organizations, larger businesses) have such IR teams, while most small and medium sized organizations create ad hoc IR teams (Shedden et al. 2010). Within organizations, the role of incident response teams is often that of 'fire-fighters' (Jaikumar 2002, Ahmad et al. 2012).

## 3.2 Role of Information Warfare Capabilities in IR

Traditionally, warfare strategies cater for multiple contingencies against the leading hypothesis of the operational plan. This approach leads to a robust response generated at the spur of the moment with the available resources in an uncertain environment (e.g. an adversary's offensive movement). Amongst the host of military strategies, IW is a strategy which is non-kinetic and aims at protecting one's own information organization while disrupting that of the adversary. The IW paradigm is predominantly response oriented. Through the concept of defence-in-depth, contingency planning and maintaining reserves at all levels, the defensive manoeuvrers are well adapted against unknown offensives of the adversaries.

IW is an alternative strategy; comprising actions which are intended to protect, exploit, corrupt, deny or destroy information or information resources to achieve a competitive advantage over an adversary (Denning & Denning 2010). Many writers have considered its utility beyond warfare as it deals with information and IS (Taddeo 2011; Armistead 2010; Hutchinson & Warren 2012). In US military doctrine, IW is the integrated employment of capabilities (core, supporting and related) to affect adversaries' information and information systems while protecting their own, thus influencing decision-making (Field Manual 3-13, 2014; Field Manual 33-1-1 2003). From the IW capabilities only two capabilities are directly applicable in corporate environments. Therefore, in this research **information warfare security capabilities** refer to: *information warfare practices incorporating capabilities of* **operational security** *(Opsec) and* **deception**.

### 3.2.1 Operational Security (Opsec)

Opsec is a process designed to meet operational needs by mitigating risks associated with specific vulnerabilities to deny adversaries critical information and observable indicators (Field Manual 3-13, 2014). It can be used to carry out analysis of adversaries, vulnerabilities, assessment of risk and application of appropriate protective measures. Its scope can be extended in IS to include monitoring and knowing about the strategic purposes of threats e.g. knowledge about the phases of APTs and objective of the intruders in each phase can be ascertained.

### 3.2.2 Deception

Deception comprises actions executed to deliberately mislead adversary's decision making about own capabilities, intentions, and operations, causing them to take specific actions (or inactions) (Field Manual 3-13, 2014). It is fundamentally about the hiding of facts and presenting the false. By incorporating deception in security mechanism, response becomes better compared to traditional security controls (Almeshekah et al. 2016). Its utility in shape of honey pots, tags, breadcrumbs is already in practice in the IS domain. The prime objective remains to adopt an approach that moves away from prevention to response by getting to know the strategies, methods, practices and objectives of the threat by interaction into safe and imitated environments through deception techniques.

## 3.3 Information Warfare Enabled Dynamic IR Capabilities (WEDIRC)

Through DC the role of important internal components of the firm are revived, changed (Zahra et al. 2006; Helfat et al. 2007). Literature categorizes these capabilities as ordinary or *substantive capabilities* (to solve ordinary or routine problems) (Winter 2003; Zahra et al. 2006) and the *dynamic* or *higher-level* capabilities; that essentially change the ordinary capabilities (Winter 2003). Teece (2018) has further elaborated the layer of DC, into "*micro-foundations*" and *higher-order capabilities*. He argues that through the micro-foundations, recombination of existing ordinary capabilities as well as the development of new capabilities is possible. Teece hypothesize that micro-foundations are guided by the high-order DC following *sensing, seizing, and transforming* phenomenon (Teece 2018).

In the same vein, our argument is that IW capabilities (response oriented) can be integrated with IR capabilities (prevention-oriented), to constitute higher order DC. By adopting IW concepts such as Opsec, the strategic purpose of APTs in its different phases can be discerned. Which is a shaping and sensing process. Subsequently by applying deception techniques such as honey pots, breadcrumbs etc; critical assets of IS can be protected. Once phases of attack and purposes in each phase have been determined, combination of deception and Opsec (e.g. kill chain) can be used to disrupt the attacker; essentially a seizing process. The whole process of this integration is the transforming phenomenon. In this paper we refer to the IW enabled dynamic IR capabilities (WEDIRC) as: **Highest level specialized dynamic capabilities resulting from integration of relevant capabilities of IW (deception, Opsec) and IR capabilities.**

## 3.4 Enterprise Security Performance

There are many criterions for gauging enterprise security performance however, scholars invariably challenge them. The security performance of an organization is a sum total of its effective control against internal and external frictions, impediments affecting its achievements of tactical and strategic goals ensuring the integrity, confidentiality and availability of its functions and critical information (Andress 2014; Naseer et al. 2016; Naseer et al. 2018). In this research we are referring to *security performance* **as the ability of the organizations to responds to unknown threats in the context of IR.**

## 4 The Proposed Research Framework

In the research framework (see Figure 1) we argue that conventional IR capabilities and the relevant capabilities, practices from the IW strategies are indirectly associated with enterprise security performance via information warfare enabled dynamic IR capabilities. When conventional IR and IW capabilities interact with each other, higher level information warfare enabled dynamic IR capabilities emerge. These enhanced high level WEDIRC then impact the overall enterprise security performance.

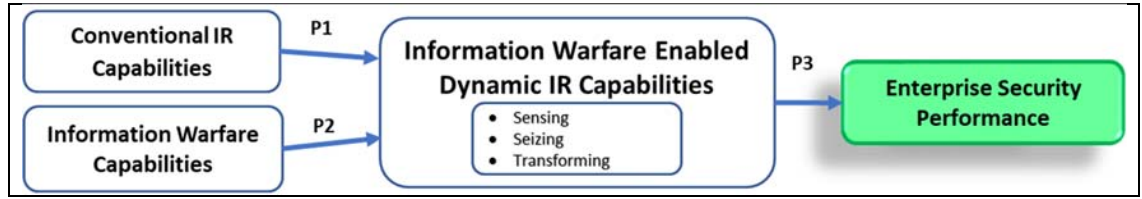

*Figure 1. Enterprise Security Performance impacts from information warfare enabled dynamic IR capabilities*

### 4.1 Conventional IR Capabilities

InfoSec practices have matured significantly over the period time, resulting in the development of many standards, controls and frameworks such as the ISO 27000. Using these standards, organizations implement their security controls. Within InfoSec practices, IR includes activities and practices which organizations use to sense and eliminate InfoSec incidents (Wiik et al. 2005; Shedden et al. 2010; Ahmad et al. 2012). Through IR process, organisations deal with attacks on their enterprise assets (Stephenson 2003). These pre-defined controls based on standard frameworks enable organizations to deal with threats that are predictable in nature (Baskerville 2014). Therefore, we propose that:

***P 1****: Firms with higher level of conventional IR capabilities will also have a higher level of WEDIRC.*

### 4.2 Information Warfare Capabilities

Since attackers are mounting well-resourced, unique, targeted and persistent attacks, the prevention-oriented philosophy with predefined controls is inadequate for the more dynamic threat environment (Baskerville 2014). With the passage of time more sophisticated and improved controls have been developed, yet the attackers are mounting even more complex and innovative purposive threats. IW security practices incorporate dynamism in threats and risks (Baskerville 2005). Through the relevant IW capabilities, a better response-oriented paradigm contributes positively in well integrated IW enabled dynamic IR capabilities (WEDIRC) to deal with unpredicted or unknown security threats. Since the response-oriented approach can assist in effectively handling new, unknown events (Baskerville 2014). So, we propose that:

***P 2****: Firms with higher level of information warfare capabilities will also have a higher level of WEDIRC.*

### 4.3 The Impact of Information Warfare Enabled Dynamic IR Capabilities (WEDIRC) on Enterprise Security Performance

Through the micro-foundations, the reconfiguration of existing ordinary capabilities as well as the development of new capabilities of the organizations is possible by following the *sensing, seizing, and transforming* phenomena (Teece 2018). Also, IR capabilities positively contribute to higher level integrated WEDIRC in the prevention domain, whereas, IW (IO) relevant capabilities have positive contribution in the response domain (Baskerville 2014). Further, we can infer that the integration of IW capabilities and conventional IR capabilities results in the highest order dynamic capabilities. These WEDIRC can ensure better knowledge management and obtain competitive advantage even under rapidly changing environment, thus enhancing the security performance. Therefore, we propose:

***P 3****: Firms with higher level of WEDIRC will also have a higher level of enterprise security performance.*

## 5 Conclusion and Future Work

Since the focus of InfoSec is on prevention, there is a mismatch between current security practices and the techniques adopted by APTs, suggesting that organisations need to develop better response capabilities. IW practices are more response oriented, so they provide a basis to enhance the response capabilities of organizations. Since military strategists already conceptualize conflict in an abstract 'battlespace', visualizing the unfolding of an adversary's capabilities therefore, their adaptation in IS can be easily materialized.

In this paper following DC theory we had argued that by adopting IW practices the InfoSec of organizations can be enhanced. We have introduced new activities for InfoSec practitioners in the domain of deception and Opsec. Therefore, this research opens venues for several useful contributions in multi-disciplinary environments of InfoSec and IW.

We intend to carry out several case studies to collect real world data through semi structured interviews to validate and refine our conceptual framework. For this purpose, interviews of security personnel (ranging from CISO level to operational IR people) from 8 to 10 organizations will be conducted. We will use scenario building, thematic and structured questions to analyse the impact of existing response-oriented practices (we will interpret these using the IW paradigm) on organizational IR. Based on the data collected hypothesis will be generated to finalize the research proceedings.